\documentclass[12pt]{article}
\usepackage{amsmath, latexsym}
\usepackage[english]{babel}      
\usepackage[T1]{fontenc}       
\begin{document}
\pagenumbering{arabic}
\begin{titlepage}
\title{Gravitational Pressure, apparent horizon and thermodynamics of FLRW universe in the
teleparallel gravity}
\author{J. F. da Rocha-Neto $\, ^{1 \dagger}$ and B. R. Morais $\, ^{1, 2 \star}$}
\date{}
\maketitle
\begin{center}
1 Instituto de F\'isica, Universidade de Bras\'ilia, \\
70910-900, Bras\'ilia, DF, Brazil.
\end{center}
\begin{center} 
2  Instituto Federal (de Educa\c c\~ao, Ci\^encia e Tecnologia)\\
de Goi\'as, 74055 - 110, Goi\^ania - GO, Brazil.
\end{center}

\begin{abstract}
In the context of the teleparallel equivalent of general relativity 
the concept of gravitational pressure and gravitational energy-momentum 
arisen in a natural way. In the case of a Friedmann-Lemaitre-Robertson-Walker 
space FLRW we obtain the total energy contained inside the apparent horizon 
and the radial pressure over the apparent horizon area. We use these 
definitions to written a thermodynamics relation $T_{A}dS_{A} = dE_{A}+P_{A}dV_{A}$ 
at the apparent  horizon, where $E_{A}$ is the total energy inside the apparent 
horizon, $V_{A}$ is the areal volume of the apparent horizon, $P_{A}$ is the 
radial pressure over the apparent horizon area, $S_{A}$ is the entropy 
which can be assumed as one quarter of the apparent horizon area only for a 
non stationary apparent horizon. We identify $T_{A}$  as the temperature at 
the surface of the apparent horizon. We shown that for all expanding 
accelerated FLRW model of universe the radial pressure is positive.

\end{abstract}
 


{\footnotesize
\noindent $\dagger$ rocha@fis.unb.br}\\
\noindent $\star$ breytmorais@gmail.com

\end{titlepage}

\section{Introduction}
\noindent

Using quantum mechanics together with general relativity Bekenstein\cite{Bekenstein} 
and Hawking\cite{Hawking} showed that black hole horizons have entropy and temperature 
associated with its horizon and that black holes behaves like a black body emitting thermal 
radiation. This discovery led to the formulation of thermodynamics of black holes with 
an entropy proportional to its horizon area and a temperature proportional to its surface 
gravity at the horizon. It is widely known in the literature that the 
Hawking temperature and horizon entropy together with the mass of the black hole, 
which is identified with its energy, satisfy the first law of  thermodynamics.  
Since the relations for black hole entropy and temperature are geometrical quantities 
associated with the black holes geometry, the thermodynamics analysis of black holes 
are in general based on geometrical aspects of black holes. In addition, it is believed 
that there exist some connection between black hole thermodynamic and Einstein's 
equations\cite{Padman}. After the discovery by Hawking that black holes emit radiation 
with a temperature proportional to its surface gravity at the event horizon\cite{Hawking}, 
Gibbons and Hawking showed that from a thermodynamic point of view the de Sitter cosmological 
event horizon have the same properties, behaving like a Schwarzchild horizon\cite{Gibbons}.
 
The initial literature on thermodynamic of black holes was focused on static and 
stationary black holes. The thermodynamic analysis of black holes evolving with 
time where we may have: black hole collapse, black hole evaporation or  black 
hole interacting with another black hole, requires the generalization of the 
concept of event horizon to dynamical horizon. A dynamical horizon define a 
marginally trapping surface on which no timelike Killing vector is available. 
With this generalization, thermodynamics properties of black holes
could be generalized to others type of horizons other than black holes horizon 
event \cite{Ashtekar, Nielson, Booth}. In fact, in the Friedman-Lemaitre-Robertson-Walker 
FLRW space  always exist a dynamical apparent horizon on which it is possible to 
formulation thermodynamics analysis (see \cite{Faraoni} and references therein). 
Recently many authors using different approaches have shown that from the first 
law of thermodynamics at the apparent horizon of FLRW space it is possible 
to derive the Friedman's equations \cite{Cai}.

For a static spherically symmetric black hole of mass $M$, the first law of 
thermodynamic on the evente horizon is in general written as $dE = TdS$. Where 
$E = M$ is the energy, $T$ the temperature and $S= A/4$ is the entropy,
$A$ is the horizon area.  Looking at the relationship that states the first
law of thermodynamics for black holes we note the absence of a work term due 
the pressure $P$ at the event horizon. The pressure $P$ can be due the source 
of matter and the gravitational field. 
Trying to solve this problem in Refs. \cite{Padman, Paranjape}, the authors 
consider a  spherically symmetric space time to show
that it is possible to derive the Einstein's
equations at the event horizon  from the thermodynamic identity $dE + PdV = TdS$.
Where in Refs. \cite{Padman, Paranjape} $P$ is strictly due the term of source in 
Einstein's equations. In addition, the term of energy present in the first law of 
thermodynamic for  black holes takes into account only the mass of the black hole, 
leaving out the contribution of energy due the gravitational field. For instance Brown 
and York\cite{York} using a quasi-local energy expression showed that the energy 
inside the event horizon of a Schwarzchild black hole is given by $2M$.  

In this work using the teleparallel equivalent of general relativity TEGR we 
obtain an equation  for the thermodynamic relation $ T_{A}dS_{A} = dE_{A} + P_{A}dV_{A}$ 
at the apparent horizon of a general FLRW space. Where $E_{A}$ represent the 
total energy (gravitational plus matter) defined in the context of the TEGR.  
The definition of gravitational radial pressure $P_{A}$ obtained in this analysis 
follows from the fields equations and from the gravitational energy-momentum four-
vector. The definition of pressure arises 
naturally from the time derivative of the spatial components of the 
energy-momentum four-vector. In the case of FLRW universes in accelerated 
expansion we will show that the radial pressure at the apparent horizon is 
positive. The infinitesimal element of volume $dV_{A}$ is obtained out of 
the areal volume of the apparent horizon. 

In the next section we present a brief summary of the Lagrangian formulation of 
the TEGR and show how we can get the definitions of the total energy-momentum 
four-vector $P^{a}$, of the gravitational energy-momentum tensor $t^{\mu\nu}$ and 
the gravitational radial pressure $P_{A}$ that naturally arise from the definition of 
the energy-momentum four-vector and the fields equations of the formalism. In Section III 
we apply these definitions to the FLRW models of space to obtain the total energy 
inside the apparent horizon as well as the radial pressure over the
apparent horizon. In Section IV we write a thermodynamic relation $T_{A}dS_{A} = 
dE_{A}+P_{A}dV_{A}$ at the apparent horizon and, for a non stationary apparent 
horizon, we obtain a expression for the temperature $T_{A}$. 
In Section V we present our final considerations.   

We use the following notation: space-time indexes $\mu$, $\nu$, ... and 
$SO(3,1)$ $a$, $b$, ... rum from $0$ to $3$. Time and spaces indexes 
are indicated as $\mu = 0, i$, $a = (0), (i)$. The tetrad 
field is denoted by $e^{a}\,_{\mu}$, and the torsion tensor 
as $T^{a}\,_{\mu\nu} = \partial\mu e^{a}\,_{\nu}
- \partial\nu e^{a}\,_{\mu}$. The flat Minkowski metric is denoted by 
$\eta_{ab} = e_{a\nu}e_{b}\,^{\mu} = (-1, 1, 1, 1)$ and $c= G = 1$. 

\section{The energy-momentum four-vector and pressure in the TEGR}
\noindent

Before we present the definitions of energy-momentum and pressure in the TEGR, 
let us present a brief summary of the teleparallel equivalent of general relativity.
In this framework the gravitational field is represented in terms of the tetrad 
field $e^{a}\,_{\mu}$, and the Lagrangian density is constructed out of a quadratic 
combination of the torsion tensor $T_{a\mu\nu} = \partial_{\mu}e_{a\nu} - 
\partial_{\nu}e_{a\mu}$ that is related to the antisymmetric part of the 
Weitzenb\"ock connection $\Gamma^{\lambda}\,_{\mu\nu} = e^{a\lambda}\partial_{\mu}e_{a\nu}$. 
To show the equivalence of the TEGR with general relativity 
constructed in the usual metric formulation, let us consider the Christoffel symbols

\begin{equation}
\mathring{\Gamma}_{\lambda\mu\nu} = {1\over 2}(\partial_{\mu}g_{\nu\lambda} 
+ \partial_{\nu}g_{\mu\lambda} - \partial_{\lambda}g_{\mu\nu}),
\label{eq.2,1}
\end{equation}
using the relation between the metric tensor and the tetrad field, $g_{\mu\nu} = e^{a}\,_{\mu}e_{a\nu}$,
it is possible to write it in terms of the Weitzenb\"ock and the torsion-free Levi-Civita connections as 
\begin{equation}
\mathring{\Gamma}_{\lambda\mu\nu} = e^{a}\,_{\lambda}(\mathring{\omega}_{\mu ab})e^{b}_{\nu} + 
e^{a\lambda}\partial_{\mu}e_{a\nu},
\label{eq.2.1a}
\end{equation}
with the Levi-Civita connection $\mathring{\omega}_{\mu ab}$  given by
\begin{equation} 
\mathring{\omega}_{\mu ab} = -K_{\mu ab}
\label{eq.2.1b}
\end{equation}
where $K_{\mu ab} =  {1\over 2}e_{a}\,^{\lambda}e_{b}\,^{\nu}(T_{\lambda\mu\nu} + 
T_{\nu\lambda\mu} - T_{\mu\nu\lambda})$ is the contorsion tensor 
and $T_{\lambda\mu\nu} = e^{a}\,_{\mu}e^{b}\,_{b}T_{\lambda ab}$. 
With the torsion-free Levi-Civita connection it is possible to write a curvature 
tensor $R^{a}\,_{b\mu\nu}(\mathring{\omega})$, and with the identity given
in Eq. (\ref{eq.2.1b}) its scalar curvature may be written identically as 
\cite{Ref21, Ref22,Ref23}  
\begin{equation}
eR(\mathring{\omega}) = -e({1\over 4}T^{abc}T_{abc} + {1\over 2}T^{abc}T_{bac} - 
T^{a}T_{a}) + 2\partial_{\mu}(eT^{\mu}),
\label{eq.1b}
\end{equation}
where $e$ is the determinant of the tetrad field.
Therefore in the context of the TEGR  the Lagrangian density for the gravitational
and matter fields is written as 
\begin{eqnarray}
L&=& -\kappa e(\frac{1}{4}T^{abc}T_{abc}+\frac{1}{2}T^{abc}T_{bac}-
T^aT_a) - \frac{1}{c}L_m\nonumber \\
&\equiv& -\kappa e\Sigma^{abc}T_{abc}-{1\over c}L_{m}\,, 
\label{2.2}
\end{eqnarray}
in which $\kappa = c^{3}/16\pi G\,$, and $\Sigma^{abc}$ is defined by
\begin{equation}
\Sigma^{abc}= \frac{1}{4} (T^{abc}+T^{bac}-T^{cab})
+\frac{1}{2}( \eta^{ac}T^b-\eta^{ab}T^c)\,,
\label{2.3}
\end{equation}
and $L_m$  is the Lagrangian density for the matter fields. The Lagrangian density 
$L$ in Eq. (\ref{2.2}) is invariant under the global SO(3,1) transformations. 
The absence of the divergence term  on the right-hand side of Eq. (\ref{2.2}) 
preclude the invariance of $L$ under arbitrary local $SO(3,1)$ transformations.

The variation of $L$ with respect to $e^{a\mu}$ gives the field equations
\begin{equation}
e_{a\lambda}e_{b\mu}\partial_\nu (e\Sigma^{b\lambda \nu} )-
e (\Sigma^{b\nu}\,_aT_{b\nu\mu}-
\frac{1}{4}e_{a\mu}T_{bcd}\Sigma^{bcd} )=\frac{1}{4\kappa}eT_{a\mu}\,,
\label{2.4}
\end{equation}
where $eT_{a\mu}=\delta L_m / \delta e^{a\mu}$. In the following we will make $c = G = 1$. 
The Eq. (\ref{2.4}) can  be rewritten as

\begin{equation}
\partial_\nu(e\Sigma^{a\lambda\nu})=\frac{1}{4\kappa}
e\, e^a\,_\mu( t^{\lambda \mu} + T^{\lambda \mu})\;,
\label{2.5}
\end{equation}
where $T^{\lambda\mu}=e_a\,^{\lambda}T^{a\mu}$ is the energy-momentum tensor 
of the matter fields and $t^{\lambda\mu}$  defined by

\begin{equation}
t^{\lambda \mu}=\kappa(4\Sigma^{bc\lambda}T_{bc}\,^\mu-
g^{\lambda \mu}\Sigma^{bcd}T_{bcd})\,.
\label{2.6}
\end{equation}
is  identified as the energy-momentum tensor of the gravitational field \cite{Ref24}.

Due the antisymmetry in the last two indexes of 
$\Sigma^{a\mu\nu}$,  it follows from Eq. (\ref{2.5}) that

\begin{equation}
\partial_\lambda
\left[e\, e^a\,_\mu( t^{\lambda \mu} + T^{\lambda \mu})\right]=0\,.
\label{2.7}
\end{equation}
Since in the last equation we have a four-divergence of a contra variant four-vector
density, using the Guass's theorem (see for instance section 7.4 
of Ref. \cite{d'Inverno}) it is possible to write a \textit{continuity 
equation}, that is
\begin{equation}
\frac{d}{dt} \int_V d^3x\,e\,e^a\,_\mu (t^{0\mu} +T^{0\mu})
=-\oint_S dS_j\,
\left[e\,e^a\,_\mu (t^{j\mu} +T^{j\mu})\right]\,,
\label{2.8}
\end{equation}
where $S$ is the surface that enclose the arbitrary volume $V$ of the thee-dimencional
space and $dS_{j}$ is given by
$$dS_{j} = {1\over 2!}\epsilon_{jkl}\,dS^{kl},\,\,\,\,\,
dS^{kl} = det\begin{pmatrix} dx^{k}&dx'^{k}\\ dx^{l}&dx'^{l}\end{pmatrix},$$
where $\epsilon_{ijk}$ is the Levi-Civita alternating symbol with $\epsilon_{123} = 1$.

On the left-hand side of Eq. (\ref{2.8}) we have the time derivative of the 
total energy-momentum four-vector $P^{a}$ 
involved by the volume $V$ of the three-dimensional space 

\begin{equation}
P^a=\int_V d^3x\,e\,e^a\,_\mu (t^{0\mu} 
+T^{0\mu}),
\label{2.9}
\end{equation}
and on the right-hand side we identify the quantities 

\begin{equation}
\Phi^{a}_{g} = \oint_{S} dS_{j}(e e^{a}\,_{\mu}t^{j\mu}),
\label{2.10}
\end{equation}

\begin{equation}
\Phi^{a}_{m} = \oint_{S} dS_{j}(e e^{a}\,_{\mu}T^{j\mu}), 
\label{2.11}
\end{equation}
as the fluxes of  energy-momentum of the gravitational field and  
matter per unit of time, respectively. $S$ represent the spatial surface
that enclose the volume $V$. From Eq. (\ref{2.8}) and the
definitions in Eqs. (\ref{2.9}), (\ref{2.10}) and (\ref{2.11}) we have
\begin{equation}
{dP^{a}\over dt} = -\Phi^{a}_{g} - \Phi^{a}_{m}.
\label{2.11a}
\end{equation}

Using Eq. (\ref{2.5}), Eq. (\ref{2.9}) may be written in terms of 
$\Pi^{ai}=-4\kappa e\,\Sigma^{a0i}$,  which is the density of momentum 
canonically conjugate to $e_{ai}$ 

\begin{equation}
P^a=-\int_V d^3x \partial_i \Pi^{ai}\quad = -\oint_{S}dS_{i}\Pi^{ai}.
\label{2.12}
\end{equation}
The definition of energy momentum four-vector $P^{a}$
in the above equation was obtained for the first time in the  context of 
Hamiltonian formulation of the TEGR in the vacuum  \cite{Ref25}. In non-empty 
space times $P^{a}$ represents the total energy momentum four-vector of the gravitational 
and matter fields. This definition is invariant under coordinate transformations 
of the three-dimensional space and under time reparametrization. 
The component $a = (0)$ of $P^{a}$ give us the total energy contained inside 
the surface $S$ that ensure the volume $V$.

Substituting Eq. (\ref{2.12}) on the left-hand side of Eq. (\ref{2.8}) and 
using Eq. (\ref{2.5}), the Eq. (\ref{2.8}) is written as

\begin{equation}
{{dP^a}\over {dt}}=
-\oint_S dS_j\,\phi^{aj},
\label{2.13}
\end{equation}
where
\begin{equation}
\phi^{aj}=4\kappa\partial_\nu(e\Sigma^{aj\nu}) = ee^{a}\,_{\mu}(t^{j\mu}+T^{j\mu})\,.
\label{2.14}
\end{equation}

If we now restrict the Lorentz index $a$ to be $a = (i)$, with $i = 1, 2, 3,$ 
the Eq. (\ref{2.13}) can be written as 
\begin{equation}
{dP^{(i)}\over dt} = - \oint_{S}dS_{j}\phi^{(i)j}.
\label{2.15}
\end{equation}
The left-hand side of Eq. (\ref{2.15}) has dimension of force and hence 
the density $\phi^{(i)j}$ on its right-hand side in dimensional 
coordinates has dimension of force per unit 
of area and represents a pressure density along the $(i)$-direction over an 
element of area oriented along the $j$-direction \cite{Ref24}.
In cartesian coordinates $j = 1, 2, 3$ represents the directions $\hat{x}, \hat{y}, \hat{z}$, 
respectively. In spherical type coordinates we fix $j = r, \theta, \phi$, as 
consequence $j = 1$ is associated with the radial direction therefore,  
to obtain the radial pressure we need consider only the index $j = 1$.
Thus in spherical type coordinates we define the radial density of force as 
\begin{equation}
-\phi^{(r)1} = -(\sin\theta\cos\phi \phi^{(1)1}+\sin\theta\sin\phi\phi^{(2)1}+\cos\theta\phi^{(3)1}),
\label{2.16}
\end{equation}
therefore, from the expression above we define the radial force as
\begin{equation}
f(r) = \int_{0}^{2\pi}d\phi\int_{0}^{\pi}d\theta[-\phi^{(r)1}].
\label{2.17}
\end{equation}
This expression was applied recently in the study of the thermodynamics of Kerr \cite{Ref26} and
Reissner-Nordstr\"om \cite{karl-jf} black holes, respectively. In Ref. \cite{Ref26}, it is shown
that the efficiency of the Penrose process for a Kerr black hole 
is lower than in the thermodynamic formulation in general relativity. While in Ref. \cite{karl-jf} 
the authors show that for a Reissner-Nordstr\"om black hole 
$TdS \geq (\kappa/8\pi)dA = TdS_{BH}$, where $S_{BH}$ is the standard Bekeinstein-Hawking entropy,
and the equality is valid only for a Schwarzchild black hole.  
In the next section we will obtain expressions for the 
energy and the radial pressure associated with the apparent horizon of the FLRW  
models of universe to write a thermodynamic relation $T_{A}dS_{A}$ entirely in the
context of the TEGR, with no {\it a priori} identification between $T_{A}dS_{A}$ and the
variation $dA_{A}$ the area of the apparent horizon.

\section{Energy and Radial Pressure in a FLRW Universe}

\noindent

In comoving coordinates $(r, t, \theta, \phi)$ the FLRW line element is given by
\begin{equation}
ds^{2} = -dt^{2}+a^{2}(t)\left\{{dr^{2}\over 1 - kr^2} + 
r^{2}\left (d\theta^{2}+ \sin ^{2}\theta d\phi^{2}\right)\right\}\,,
\label{2.18} 
\end{equation}
where $k =0, 1, -1$ is the curvature index and $a(t)$ is the scale factor. 
If the FLRW model of universe contains a perfect fluid with energy-momentum tensor
\begin{equation}
T_{\mu\nu}= pg_{\mu\nu}+\left(\rho + p\right)u_{\mu}u_{\nu}\,, 
\label{2.19} 
\end{equation}
where $\rho$, $p$ and $u^{\mu}$ are the energy density, pressure and the four-velocity 
field of the fluid, respectively. The Einstein's equations read as
\begin{equation}
R_{\mu\nu}-\frac{1}{2}Rg_{\mu\nu}  = 8\pi T_{\mu\nu}. 
\label{2.20} 
\end{equation}
From the two equations above one has
\begin{equation}
2\frac{\ddot{a}}{a}+\frac{\dot{a}^2}{a^{2}}+\frac	{k}{a^2} =-8\pi p,
\label{2.21}
\end{equation}

\begin{equation}
\frac{\dot{a}^2}{a^{2}}+\frac{k}{a^2} = \frac{8\pi\rho}{3},
\label{2.22}
\end{equation}
here the over dot represent the derivative with respect to the cosmological time $t$.
An important equation following from the two above equations is 
\begin{equation}
{\dot{a}^{2}\over a^{2}} + {k\over a^{2}} - {\ddot{a}\over a} = 4\pi(\rho + p),
\label{2.22a}
\end{equation}

The dynamical apparent horizon of the FLRW universe is determined by the relation
$h^{AB}\partial_{A}R(r,t)\partial_{B}R(r,t) = 0$, where $R(r,t) = a(t)r$ is the 
areal radius and $h_{AB} = [-1, a^{2}(t)/(1-kr^2)]$ represent the transverse
two metric spanned by $x^{A} = (t,r)$ where the indexes $A$ and $B$ can take 
the values $(0,1)$  \cite{Faraoni}. 
This condition implies that  the gradient $\nabla R(r,t)$ is a null vector 
on the surface of the apparent horizon. Out of the explicit evaluation of  apparent 
horizon for the FLRW universe we obtain the apparent horizon radius as
\begin{equation}
R_{A} = {1\over \sqrt{H^{2} + k/a^{2}}}, 
\label{2.23}
\end{equation}
where $H = \dot {a}/a$ is the Hubble parameter. Note also that due the Eq. (\ref{2.22}), 
the argument of the square root in (\ref{2.23}) is positive for positive densities $\rho$ what
ensures that the apparent horizon radius is real. 

In order to obtain the tetrad field related to a metric tensor we consider the relation 
$e^{a}\,_{\mu}e_{a\nu} = g_{\mu\nu}$. To analyze the physical properties of a metric tensor, 
we chose a tetrad field adapted to a field of static observers, whose trajectories and velocities
in space-time are given by $x^{\mu}(s)$ and $u^{\mu}(s) = dx^{\mu}/ds$, respectively. Where
$s$ is the proper time and we identify $u^{\mu} = e_{(0)}\,^{\mu}$. A set of tetrad field adapted
to static observers is achieved by imposing the following conditions: (1) $e_{(0)}\,^{i} = 0$, which
implies that the three spatial translational velocities of the adapted observer are zero; 
(2) $e_{(0)i} = 0$ which implies that the three spatial axes of the adapted observer are not rotating
with respect to a non rotating frame \cite{Ref26}.    

A set of tetrad field related to the line element given in (\ref{2.18}) and satisfying
the conditions (1) and (2) above is given by
\footnote{If we define the tetrad field as $e^{a}\,_{\mu} = {\partial y^{a}\over \partial x^{\mu}}$, 
we can immediately shows that all the components of the torsion tensor will be nulls
i.e, $T^{a}\,_{\mu\nu} = 0$. In the context of the TEGR the gravitational field is described 
by configuration of $e^{a}\,_{\mu}$ such that $T_{a\mu\nu} \neq 0$. 
In addition if we consider a tetrad field such as that given in Eq. (14) of the Ref. \cite{Ulhoa}, 
some drawbacks may appear. For example, the emergence of non-vanishing torsion components
in Minkowisk space-time. These drawbacks may be eliminated by means of the regularization
procedures presented in Ref.\cite{JWJF}.} 

\begin{equation}
e_{a\mu} = \left(\begin{array}{cccc}
-1 & 0 & 0 & 0\\
0  & \alpha\sin\theta\cos\phi & ar\cos\theta\cos\phi & - ar\sin\theta\sin\phi \\
0  & \alpha\sin\theta\sin\phi & ar\cos\theta\sin\phi & ar\sin\theta\cos\phi \\
0  & \alpha\cos\theta         & - ar\sin\theta       & 0
\label{2.23a}
\end{array}
\right),
\end{equation}
where 
\begin{equation}
\alpha = {a(t)\over \sqrt{1 - kr^2}},
\label{2.24}
\end{equation}
and its determinant is $e = \alpha a^{2}r^{2}\sin\theta$. We emphasize that
due to symmetry in the line element in Eq.(\ref{2.18}), it is more convenient
to write the tetrad field $e_{a\mu}$ in spherical coordinates. 

We are now in position to evaluate all necessary quantities to obtain the energy 
contained inside the apparent horizon and the radial pressure over the apparent horizon.
We emphasize that only after doing all the calculations is that we apply the results on the
surface of apparent horizon. 
The calculations are lengthy, but otherwise straightforward. Therefore we find important 
to show some intermediate steps. First we need to calculate the necessary quantities 
$\Sigma^{\alpha\mu\nu}$ related to Eq. (\ref{2.3}). Using the definition of torsion tensor
$T_{a\mu\nu} = \partial_{\mu}e_{a\nu} - \partial_{\nu}e_{a\mu}$, they are given by
\begin{eqnarray}
\Sigma^{001} &=& {1\over ra\alpha^{2}}(\alpha - a),\nonumber \\
\Sigma^{110} &=& -{\dot{a}\over a \alpha^{2}}, \nonumber \\
\Sigma^{220} &=& -{\dot{a}\over r^{2}a^{3}}, \nonumber \\
\Sigma^{330} &=& -{\dot{a}\over a^{3}r^{2}\sin^{2}\theta}, \nonumber \\
\Sigma^{212} &=& {(\alpha - a)\over 2r^{3}a^{3}\alpha^{2}}, \nonumber \\
\Sigma^{313} &=& {(\alpha -a)\over 2r^{3}a^{3}\alpha^{2}\sin^{2}\theta}. \nonumber \\
\label{2.25}
\end{eqnarray}
All the others components of $\Sigma^{\alpha\mu\nu}$ are zero. Since  $e^{(0)}\,_{i} = 0$ 
and $e^{(0)}\,_{0} = 1$, we have that 
$\Sigma^{(0)01} = e^{(0)}\,_{0}\Sigma^{001} = \Sigma^{001}$.
 Finally, from the four-divergence in Eq. (\ref{2.14}) and using the 
results presented in Eq. (\ref{2.25}), the quantities $\phi^{(i)1}$
are given by
\begin{eqnarray}
\phi^{(1)1} &=& -4\kappa\left[\partial_{0}(\dot{a}a){r^{2}} + 1 - 
\sqrt{1 - kr^{2}}\right]\sin^{2}\theta \cos\phi,\nonumber \\
\phi^{(2)1} &=& -4\kappa\left[\partial_{0}(\dot{a}a){r^{2}} + 1 - 
\sqrt{1 - kr^{2}}\right]\sin^{2}\theta \sin\phi, \nonumber \\
\phi^{(3)1} &=& -4\kappa\left[\partial_{0}(\dot{a}a){r^{2}} + 1 - 
\sqrt{1 - kr^{2}}\right]\sin\theta \cos\theta.
\label{2.26}
\end{eqnarray}
Note for example that from Eq. (\ref{2.14}) to calculate $\phi^{(1)1}$ we should start from 
$$\phi^{(1)1} = 4\kappa[\partial_{0}(ee^{(1)}\,_{1}\Sigma^{110})+\partial_{2}(ee^{(1)}\,_{2}\Sigma^{212})+\partial_{3}(ee^{(1)}\,_{3}\Sigma^{313})],$$
where we use that $\Sigma^{\alpha\mu\mu} = 0$. The other quantities $\phi^{(i)1}$ are obtained similarly.
Now with help of Eqs. (\ref{2.26}) and after some calculations we arrive at an expression
for $\phi^{(r)1}$ given by Eq. (\ref{2.16}), it is
\begin{equation}
-\phi^{(r)1} = 4\kappa\left[\partial_{0}(\dot{a}a){r^{2}} + (1 - \sqrt{1 - kr^{2}})\right]\sin\theta,
\label{2.27}
\end{equation}
this expression will be considered later.

Let us now evaluation the energy contained inside the apparent horizon.
Remember that the density of energy is given by $\Pi^{(0)1} = -4\kappa e \Sigma^{(0)01},$ from the 
first equation in (\ref{2.25}) and Eq. (\ref{2.12}), the energy enclosed by a spherical surface of
radius $r$ is given by
\begin{equation}
E \equiv P^{(0)} = \int^{2\pi}_{0}d\phi\int^{\pi}_{0}d\theta\Pi^{(0)1} = ar (1 - \sqrt{1 - kr^2}),
\label{2.28}
\end{equation}
where we have used $\kappa = 1/16\pi$. Now if we assume $r =r'$ such that $R_{A} = a(t)r'$, and 
with the definition of $R_{A}$ in Eq. (\ref{2.23}), the energy contained inside the apparent 
horizon is given by
\begin{equation}
E_{A} = R_{A}(1 - \sqrt{1 - kR_{A}^{2}/a^{2}}).
\label{2.29}
\end{equation}
With Eqs. (\ref{2.22}) and (\ref{2.23}), $E_{A}$ may be rewritten as
\begin{equation}
E_{A} = 2M_{A} - HR_{A}^2,
\label{2.30}
\end{equation}
where 
$$M_{A} = {4\pi R_{A}^{3}\over 3}\rho,$$
is the Minsner-Sharp-Hernandez mass \cite{Faraoni}, 
which is defined only for spherically symmetric space-times.  Here we would like to remind
the readers that in Eq. (\ref{2.29}), the energy $E_{A}$ inside the apparent horizon, is not
only the Misner-Sharp-Hernandez mass. The reason is that $E_{A}$ represent the total 
energy of matter plus the energy of the gravitational field, respectively.   
From the Eq. (\ref{2.29}), we conclude that $E_{A}$ is positive for $k = 1$, 
negative for $k = -1$ and zero for $k = 0$, respectively. 
In this latter case, the negative binding gravitational energy and the positive 
matter energy inside the apparent horizon exactly cancel out resulting in an universe 
that has flat  spatial section \cite{James}.  

To obtain the radial pressure over the apparent horizon, firstly we inserting $-\phi^{(r)1}$
given by Eq. (\ref{2.27}) into Eq. (\ref{2.17}) and making the integration of the angular
variables. After the angular integration we obtain
\begin{equation}
f(r) = [\partial_{0}(\dot{a}a)r^{2} + (1 - \sqrt{1 - kr^2})].
\label{2.31}
\end{equation}
Again assuming $r = r'$ such that $R_{A} = a(t)r'$, and with the definition of $R_{A}$ 
in Eq. (\ref{2.23}), the radial force over the surface of the apparent horizon read
\begin{equation}
f_{A} = \left[\partial_{0}(\dot{a}a){R_{A}^{2}\over a^{2}} + 1 -R_{A}H\right].
\label{2.31a} 
\end{equation}
Since the FLRW models of universe are homogeneous and isotropic,  
dividing $f_{A}$ by the area of the apparent horizon we obtain the radial pressure 
over the surface of the apparent horizon.  It is given by
\begin{equation}
P_{A} = {1\over 4\pi R_{A}^{2}}\left[\left({\ddot{a}\over a} + 
{\dot{a}^{2}\over a^{2}}\right)R_{A}^{2} + 1 - R_{A}H\right].
\label{2.31b}
\end{equation}
Here we note that according Eqs. (\ref{2.10}) and (\ref{2.11}), $P_{A}$ is due the 
fluxes of gravitational field and matter, respectively. 
Taking into consideration only accelerated expanding FLRW models of universe ($\ddot{a} > 0)$, 
using the definition of $R_{A}$, for $k = 1$ or $k = 0$, $1 - HR_{A} \geq 0$ and for
$k = -1$, $H^{2}R_{A}^2 + 1 -HR_{A} > 0$, so it is not difficult to see that the radial pressure 
over the apparent horizon is positive, directed outward over it, like a tension. Since
$H^{2}R^{2}_{A} + 1 - HR_{A} > 0$ for all models of universe, the radial pressure in (\ref{2.31b})
is negative only for decelerated models of universe with 
$\ddot{a}R^{2}_{A}/a < -(H^{2}R^{2}_{A} + 1 - HR_{A})$ and in this case the pressure  
produces a force on the surface of the apparent horizon that points on the outside to the inside.
We note also that even for flat ($k = 0$) FLRW model
of universe where $E_{A} = 0$, the radial pressure over the surface of the apparent horizon is not 
necessarily zero. 

Before we close this section let us evaluate the consistency of Eq. (\ref{2.13}). Since
in this case $e_{(0)i} = e_{(i)0} = 0$, for $a = (0)$ only $\phi^{(0)1} = -4\kappa\partial_{0}
(e\Sigma^{(0)01})$ in Eq. (\ref{2.14}) is not zero so from Eq. (\ref{2.13}), at the apparent
horizon, we have 
\begin{equation}
\dot{P}^{(0)}_{A} = -\int_{0}^{2\pi}d\phi\int_{0}^{\pi}d\theta\phi^{(0)1}(R_{A}) = 
\partial_{0}\int_{0}^{2\pi}d\phi \int_{0}^{\pi}d\theta[4\kappa \Sigma^{(0)01}(R_{A})] 
= \partial_{0}E_{A}\,.
\label{3.32}
\end{equation}
To obtain the radial quantities $\Phi_{g}^{(r)}$ and $\Phi^{(r)}_{m}$ in 
Eqs. (\ref{2.10}) and (\ref{2.11}) on the surface of a sphere of 
radius $r$, we note that the radial unit vector $\hat{r}$ 
is given in term of the component of the tetrad field $e_{(i)1}$, i.e $\hat{r} = {e_{(i)1}/\alpha}$, 
so projecting the Eq. (\ref{2.14}) along of the radial direction we have
$$\phi^{(r)1} = {1\over \alpha}e_{(i)1}\phi^{(i)1} = {1\over 4\kappa\alpha}e\,g_{11}(t^{11} + T^{11}) = 
\phi^{(r)1}_{g} + \phi^{(r)1}_{m},$$
where from Eq. (\ref{2.6}) and using the quantities in Eq. (\ref{2.25}) after some calculations we obtain
$$\phi^{(r)1}_{g} = 2\kappa \sin{\theta}\left[-(Hra)^{2} - \left(2(1 - \sqrt{1 - kr^{2}}) - kr^{2}\right)\right]\,,$$
and 
$$\phi^{(r)1}_{m} = pr^{2}a^{2}\sin{\theta}\,.$$
Now performing the integrating of the two quantities above as in (\ref{2.10}) and (\ref{2.11}) on the 
surface of the apparent horizon ($r = r' = R_{A}/a(t)$), we obtain firstly the radial flow of momentum
per unit time due the matter
$$-\Phi^{(r)}_{m}(R_{A}) = -\int_{0}^{2\pi}d\phi \int^{\pi}_{0}d\theta\phi^{(r)1}_{m} = -4\pi pR^{2}_{A}\,,$$
and using the definition of $R_{A}$ in (\ref{2.23}) it is not difficult to shown that the radial flow per
unit of time due the gravitational field is given by
$$-\Phi^{(r)}_{g}(R_{A}) = -\int_{0}^{2\pi}d\phi \int^{\pi}_{0}d\theta\phi^{(r)1}_{g} 
= \left(H^{2}R^{2}_{A} + {1\over 2} - HR_{A}\right)\,.$$
To our surprise it is always positive for any model of universe expanding or contracting.  
Using the Eq. (\ref{2.21}), we have that $-4\pi p = \ddot{a}/a + 1/(2R^{2}_{A})$ therefore, from the
two equations above the total radial flow of momentum per unit time can be written as 
\begin{equation}
\dot{P}^{(r)}_{A} = -\Phi^{(r)}_{g}(R_{A})-\Phi^{(r)}_{m}(R_{A}) = \left[\left({\ddot{a}\over a} 
+ H^{2}\right)R^{2}_{A} + 1 -R_{A}H\right] = f_{A}\,,
\label{3.33}
\end{equation}
that is the same result obtained in (\ref{2.31a}).

\section{Thermodynamic of apparent horizon}

\noindent

In this section, we will use the definitions of energy and radial pressure obtained in 
the latter section to write the thermodynamic relation 
$T_{A}dS_{A} = dE_{A} + P_{A}dV_{A}.$ In order we identify $T_{A}$ with the temperature at
apparent horizon. $S_{A}$ and $V_{A}$ are the entropy and the areal volume of 
the apparent horizon, respectively. To compare our results with those obtained in the context
of general relativity, in the variation $dE_{A}$ we consider that in a infinitesimal 
interval of cosmological time $dt$, we have $dR_{A} = \dot{R}_{A}dt$ and $dH = \dot{H}dt$. 
Where here we are assuming that $\dot{R}_{A} \neq 0$, the case $\dot{R}_{A} = 0$ (a stationary 
apparent horizon) will be analyzed in the end of this section.
From Eq. (\ref{2.29}) we have
\begin{equation}
dE_{A} = \left(1 - 2HR_{A} - {\dot{H}R_{A}^{2}\over \dot{R}_{A}}\right){dA_{A}\over 8\pi R_{A}},
\label{2.32}
\end{equation}
where $A_{A} = 4\pi R_{A}^{2}$ is the areal area of the apparent horizon and $\dot{R}_{A} = 
R_{A}^{3}H(H^{2} + k/a^{2} - \ddot{a}/a)$. 
The work term due the radial pressure is given by
\begin{equation}
P_{A}dV_{A} = \left[\left({\ddot{a}\over a} + H^{2}\right)R_{A}^{2} 
+ 1- HR_{A}\right]{dA_{A}\over 8\pi R_{A}},
\label{2.33}
\end{equation}
where $V_{A} = 4\pi R_{A}^{3}/3$ is the spherical volume defined by $R_{A}$.

We are now in position to write an expression for the first law of thermodynamics, i.e

\begin{equation}
T_{A}dS_{A} = dE_{A} + P_{A}dV_{A},
\label{2.34}
\end{equation}
With the help of Eq. (\ref{2.23}) the right side of this equation can be 
simplified and we obtain
\begin{equation}
T_{A}dS_{A} = {1\over 8\pi R_{A}}\left[-2\kappa'R_{A} + (1 - HR_{A})^{2} - 
\left(HR_{A} + {\dot{H}R_{A}^{2}\over \dot{R}_{A}}\right)\right]dA_{A},
\label{2.35}
\end{equation}
where
\begin{equation}
\kappa' = {1\over 2}\Box_{(h)}R = -{R_{A}\over 2}\left({\ddot{a}\over a} + H^{2} 
+ {k\over a^{2}}\right),
\label{2.36}
\end{equation}
is the Kodama-Hayward surface gravity\cite{Ref27} and $\Box_{(h)}R$ is the d'Lambertian
of the areal radius $R(t,r) = a(t)r$ 
\begin{equation} 
\Box_{(h)}R = {1\over 2\sqrt{-h}}\partial_{A}[\sqrt{-h}h^{AB}\partial_{B}(a(t)r],
\label{2.36a}
\end{equation}
where $h_{AB} = [-1, a^{2}(t)/(1 - kr^{2})]$.
In the expression (\ref{2.35}) the last term in parentheses can be simplified and
written as
$$ - {k\ddot{a}\over \dot{a}}{R_{A}\over (\dot{a}^{2} + k - a\ddot{a})},$$
and $T_{A}dS_{A}$ can be rewritten in terms of the variation of the
Bekeinstein-Hawking entropy, i.e $dS_{BH} = dA_{A}/4$
\begin{equation}
T_{A}dS_{A} = {1\over 2\pi R_{A}}\left[-2\kappa'R_{A} + (1 - HR_{A})^{2} - 
{k\ddot{a}\over \dot{a}}{R_{A}\over (\dot{a}^{2} + k - a\ddot{a})}\right]dS_{BH}.
\label{2.37}
\end{equation}

The first term in $T_{A}dS_{A}$ is exactly twice the Kodamma-Hayward temperature at
the apparent horizon, which is given by
\begin{equation}
T_{KH} = -{\kappa'\over 2\pi} = {R_{A}\over 4\pi}\left({\ddot{a}\over a} + H^{2} + {k\over a^{2}}\right).
\label{2.38}
\end{equation}
The Kodama-Hayward temperature written as, $T_{KH} = |\kappa'|/2\pi$, is usually employed in the 
literature to write the first law of thermodynamics  at the apparent horizon of the 
FLRW model of universe. However, as pointed out in Refs. \cite{Nielsen1, Nielsen2}, 
there are several nonequivalent prescriptions for  $\kappa'$ and as consequence there 
will be several nonequivalent expressions for $T_{KH}$. 
In the following we summarize the 
behavior of $T_{A}$ for $k = 0, -1, 1$ when we assume that $S_{A} = A_{A}/4 = S_{BH}$. 

1) If we consider a flat model of universe $k = 0$ ($E_{A} = 0 = 1 - HR_{A}$), the temperature at the 
apparent horizon reduce to twice the Kodamma-Hayward temperature and is positive if 
and only if the Ricci scalar is, which correspond to equations of state satisfying 
$\rho > 3p$, and is negative if and only if the Ricci scalar is, this is a result
similar to that obtained in Ref. \cite{Faraoni} in the context of general relativity. 
Note also that  if we assume $R_{A} > 0$, the Ricci scalar in Eq. (\ref{2.36}) is negative 
if and only if,  $\ddot{a}R^{2}_{A}/a < -1$, that correspond to decelerated universe with a
radial pressure $P_{A}$ negative. If the Ricci scalar vanishes we have 
a {\it cold apparent horizon} with $T_{A} = 0$. Note that in this case
$P_{A} = 0$ and $E_{A} = 0$. The reason of $T_{A} = 2T_{KH}$ 
for $k = 0$, can be related to the fact that in the context of the TEGR 
the radial pressure over the apparent  horizon is due the fluxes of matter and 
gravitational field such that the two contributions became somehow equal, giving
twice $T_{KH}$, one from matter the other from gravity.

2) In the case of an open, $k = -1$, FLRW model of universe Eq. (\ref{2.37}) can be written as
\begin{equation}
T_{A} = {1\over 2\pi R_{A}}\left[R^{2}_{A}{\ddot a\over a} + (1 - HR_{A})^{2} + 1
+{\ddot{a} a\over \dot{a}\sqrt{\dot{a}^{2} - 1}}\,{1\over (\dot{a}^{2} - 1 - \ddot{a}a)}\right],
\label{2.39}
\end{equation}
where here we use the definition of $\kappa'$ in Eq. (\ref{2.36}). Since in this case
$\dot{a} \geq 1$, for an accelerated expanding universe $\ddot{a} \geq 0$, without violation 
the weak energy condition, $\rho + p > 0 \,\, (\dot{a}^{2} - 1 - \ddot{a}a > 0)$, the 
temperature $T_{A}$ above is positive. 

3) For a closed model of universe, $k = 1$, the situation is more complicated and  
unfortunately we do not have a general physical criterion to show that $T_{A}$ is positive.  
However, in this case with the help of the definition of $\kappa'$,
in Eq. (\ref{2.36}) the temperature $T_{A}$ can be written as
\begin{equation}
T_{A} = {1\over 2\pi R_{A}}\left[R^{2}_{A}{\ddot a\over a} + (1 - HR_{A})^{2} + 1
-{\ddot{a} a\over \dot{a}\sqrt{\dot{a}^{2} + 1}}\,{1\over (\dot{a}^{2} + 1 - \ddot{a}a)}\right].
\label{2.40}
\end{equation}

If we assume the weak energy condition as $\rho + p \geq 0$, the boundary $\rho + p = 0$
with $H \neq 0$ correspond to a stationary apparent horizon i.e 
\begin{equation}
\dot{R}_{A} = HR^{3}_{A}\left(H^{2} +
{k\over a^{2}} - {\ddot{a}\over a}\right) = 4\pi HR^{3}_{A}(\rho + p) = 0,
\label{2.41}
\end{equation}
in this case the work term $P_{A}dV_{A}$ vanishes in Eq. (\ref{2.34}). For a stationary 
apparent horizon from Eq. (\ref{2.30}) the time evolution of the energy 
inside the apparent horizon is given by $\dot{E}_{A} = -(kR^{2}_{A})/a^{2}$  
and from Eq. (\ref{2.34}) follows that
\begin{equation}
T_{A}\dot{S}_{A} = -k{R^{2}_{A}\over a^{2}}.
\label{2.42}
\end{equation}
The equation above is an important result of the paper. 
This equation states that: if, $k = 0$, the time evolution of $E_{A}\; (E_A = 0)$ 
inside the apparent horizon is zero and $T_{A}\dot{S}_{A} = 0$, a result
consistent with the fact that in this case $dE_{A} = P_{A}dV_{A} = 0$. If $k = -1$,
$E_{A}$ inside the apparent horizon is negative and increase with $t$ and
$T_{A}\dot{S}_{A} > 0$.
For a closed, $k=1$, model of universe, the result in equation above
may seem unphysical however, in this case $\dot{E}_{A}$ is always 
negative and physically this means that there is a flow of energy from the 
inside to the outside of the apparent horizon and according to the 
second law of thermodynamics it will result in a decrease of 
entropy $S_{A}$ i.e, $\dot{S}_{A} < 0$, and so the result shown in Eq. (\ref{2.42})
is consistent whenever $T_{A} > 0$. 
In this sense the apparent horizon is not an thermodynamically isolated system. 
Note that for a stationary apparent horizon with $H \neq 0$, from 
Eqs. (\ref{2.21}), (\ref{2.22}) and (\ref{2.22a})
we have $\rho = -p = constant$, $\ddot{a}/a = 1/R^{2}_{A} = constant > 0$. 
Therefore it is not obvious {\it a priori} that a constant entropy for stationary 
apparent horizon with, $\rho =-p$, $\ddot{a}/a$ and $R_{A}$ constants, 
should occur when $a$ is varying with cosmological time $t$.

For example, if we consider a cosmological constant $\Lambda > 0$ as the only source
of gravity ($\rho = p = 0$), the solution of FLRW equations in the presence
of the cosmological constant for $k = -1$ give the scale factor
$$a(t) = a_{0}\sinh(t/a_{0}),$$
and the apparent horizon radius has the constant value $R_{A} = \sqrt{3/\Lambda} = 
a_{0}$, and Eq. (\ref{2.42}) give us
\begin{equation}
T_{A}\dot{S}_{A} = {R^{2}_{A}\over a^{2}(t)}.
\label{2.43}
\end{equation}
For $k = 1$ with a cosmological constant $\Lambda > 0$ as the only source of gravity 
($\rho = p = 0$), the scale factor is given by
$$a(t) = a_{0}\cosh(t/a_{0}),$$
again the constant apparent horizon radius is $R_{A} = \sqrt{3/\Lambda} = a_{0}$, and 
from Eq. (\ref{2.42}) we have
\begin{equation}
T_{A}\dot{S}_{A} = -{R^{2}_{A}\over a^{2}(t)}.
\label{2.44}
\end{equation}
The results in Eqs. (\ref{2.43}) and (\ref{2.44}) are physically equivalent to those 
given just in Eq. (\ref{2.42}) when $\rho + p = 0$, $k = -1$ and $k = 1$, respectively.
The results presented in Eqs. (\ref{2.43}) and (\ref{2.44}) imply that in the context
of the TEGR, in general we cannot write the entropy as proportional to the area of the
apparent horizon. 

\section{Conclusions}
\noindent 

In this work we have presented the definitions of energy contained within an arbitrary 
volume enclosed by an arbitrary surface as well as the total pressure (due the fluxes of matter and 
gravitational field) on the surface in question. These definitions arise in a natural way in the
context of the teleparallel equivalent of general relativity. Considering a FLRW model of universe, we
calculation the total energy within the apparent horizon and shown that it is negative, positive
or zero for $k = -1, 1, 0$, respectively. The energy $E_{A}$ is different of Misner-Sharp-Ernandez 
mass which is given by $M_{A} = (4\pi/3)\rho R^{3}_{A} = R_{A}/2$. This difference is 
not surprise because in the TEGR the total energy enclosed by the apparent horizon 
is ascribed to the energy due the matter and others possible form of energy.   
This difference implies  that in general we cannot write the variation $dE_{A}$ 
as being proportional to the variation $dA_{A}$ the area of the apparent horizon, 
and as consequence this has implications on the thermodynamic relation for 
the apparent horizon i.e, in general we cannot write $dE_{A} + P_{A}dV_{A}$ 
as proportional to $dA_{A}$.

We also computation the total radial pressure on the surface of apparent 
horizon and show that this is not only  the pressure due the fluid inside the apparent 
horizon. The pressure computed in the context of the TEGR is due the fluxes of matter 
and gravitational field, respectively. The radial pressure $P_{A}$ given in Eq. (\ref{2.32}) is  
positive whenever $\ddot{a} > 0$. In particular for a flat model of universe, $1 - HR_{A} = 0$,
and with the equations (\ref{2.21}) and (\ref{2.22}), the radial pressure $P_{A}$ can
be written as $P_{A} = (\rho - 3p)/3$. Therefore, for a flat model of universe with
$\rho > 3p$, assuming that the density of the fluid is positive, the lower is the 
pressure of the fluid higher is the radial pressure $P_{A}$. The especial case $k = 0$
of non relativistic matter dominated universe is modeled by dust approximation with a 
pressureless matter $p = 0$, and in this case the radial pressure can be written as, 
$P_{A} = \rho/3 = (1/4\pi a)[\ddot{a} + \dot{a}^{2}/a] > 0,$ therefore the positive term, 
$[\ddot{a} + \dot{a}^{2}/a]$ appear in $P_{A}$ as an effective acceleration, which may
be responsible for the accelerated expansion of the universe. 

With the expressions of energy and pressure for the apparent horizon of FLRW space, 
for a non stationary apparent horizon, we have obtained a thermodynamic relation 
$T_{A}dS_{A} = dE_{A} + P_{A}dV_{A}$ (see Eq. (\ref{2.37})) entirely within the framework 
of the TEGR  without {\it a priori} identify $dS_{A}$ with the variation $dA_{A}$ of 
the area of the apparent horizon. To compare our result obtained in the framework of
the TEGR given in Eq. (\ref{2.35}) 
with the standard result $T_{KH}dS_{BH}$, we have rewritten the right-hand side of 
Eq. (\ref{2.35}) in terms of $dS_{BH} = dA_{A}/4$. The result is presented in the
Eq. (\ref{2.37}) which implies that in the context of the TEGR
$T_{A}dS_{A} \ne T_{KH}dS_{BH}$. In particular for $k = 0$ and $k =-1$ models of universe
that no violation the week energy condition $\rho + p > 0$ and
in accelerated expansion, $T_{A}dS_{A} > T_{KH}dS_{BH}$, which implies that if
we assume $dS_{A} = dS_{BH}$, the temperature obtained in the framework of
the TEGR is greater than the Kodama-Hayward temperature, i.e $T_{A} > T_{KH}$.
The reason that the thermodynamic relation given by Eq. (\ref{2.37}), obtained
in the framework of the TEGR is different from that obtained in the context
of general relativity is because the energy $E_{A}$ take into account the 
energy of matter and the gravitational field and that the pressure $P_{A}$ is due
the fluxes of matter and gravitational field, respectively.  

Finally, in the case of a closed, ($k = 1)$, model of universe when we assume $\dot{R}_{A} = 0$, 
(a stationary apparent horizon), we show that $\dot{E}_{A} < 0$. This means that 
there are a flow of energy from inside to outside of the apparent horizon and 
as consequence the entropy $S_{A}$ always decreases with the cosmological time $t$
which implies that the relationship in Eqs. (\ref{2.42}) and (\ref{2.44}) 
do not violate the second law of thermodynamics whenever $T_{A} > 0$. So, 
in the framework of the TEGR the apparent horizon does not constitute 
a thermodynamically isolated system.


\begin{thebibliography}{10}
\bibitem{Bekenstein}
J. D. Bekenstein, Phys. Rev. D {\bf 7}, 2333 (1973).
\bibitem{Hawking}
S. W. Hawking, Commun. Math. Phys. {\bf 43}, 199 (1975)
\bibitem{Padman} T. Padmanabhan, Class. and Quamt. Grav. {\bf 19}, 5387 (2002).
\bibitem{Gibbons}
W. Gibbons and S. W. Hawking, Phys. Rev. D {\bf 15}, 2738 (1977).
\bibitem{Ashtekar}
A. Ashtekar and B. Krishman, Living Rev. Relativity {\bf 7}, 10 (2004).
\bibitem{Nielson}
A. B. Nielson, Gene. Relativ. Gravit. {bf 41} 1539 (2009).
\bibitem{Booth}
I. Booth, Can. J. Phys. {\bf 83}, 1073 (2005).
\bibitem{Faraoni}
Valerio Faraoni, Phis. Rev. D {\bf 84}, 024003 (2011).
\bibitem{Cai}
Rong-Gen Cai and Sang Pyo Kim, J. High Energy Phys. 02 (2005);\\
M. Akbar and Rong-Gen Cai, Phys. Rev. D {\bf 75} 084003 (2007).
\bibitem{Paranjape}
Assem Paranjape, Sudipta Sarkar and P. Padmanabhan, Phys. Rev. D {\bf 74}, 104015 (2006).
\bibitem{York}
J. D. Brown and J. W. York, Phys. Rev. D {47}, 1407 (1993).
\bibitem{JWJF} 
J. W. Maluf, M. V. O. Veiga, J. F. da Rocha-Neto, Gen. Relativ. Gravit. {\bf 39}, 227 (2007). 
\bibitem{Ref21}
F. W Hehl,{\it in Proceedings of the 6th Scool of Cosmology and Gravitation on Spin, Torsion, Rotation
and Supergravity, Erice, 1979,} edited by P. G. Bergmann and V. de Sabbata (Plenum, New York, 1980).
\bibitem{Ref22}
F. W Hehl, J. D. McCrea, E. W. Mielke, and Y. Ne'eman Phys. Rep. {\bf 258}, 1 (1995).
\bibitem{Ref23}
M. Blagojevic, {\it Gravitation and Gouge Symmetries} (IOP. Bristol 2002).
\bibitem{Ref24} J. W. Maluf, Ann. Phys. (Berlin) {\bf 14}, 723 (2005).
\bibitem{d'Inverno}
Ray d'Inverno, {\it Introducing Eistein's Relativity}. Clarendon Press, Oxford, (1996).
\bibitem{Ref25} J. W. Maluf and J. F. da Rocha-Neto, Phys. Rev. D {\bf 64}, 084014 (2001).
\bibitem{Ref26} J. W. Maluf, S. C. Ulhoa and J. F. da Rocha-Neto, Phys. Rev. D {\bf 85}, 044050 (2012).
\bibitem{karl-jf}
K. H. C. Castello-Branco and J. F. da Rocha-Neto, Phys. Rev. D {\bf 88}, 024045 (2013).
\bibitem{Ulhoa} J. G. Silva, A. F. Santos, S. C. Ulhoa, The European Physical Journal C {\bf 76}, 167 (2016).
\bibitem{James} J. M. Nester, L. L. So and T. Vargas, Phys. Rev. D {\bf 78}, 044035 (2008).
\bibitem{Ref27} S. A. Hayward, Classical Quantum Gravity {\bf 15}, 3147 (1998).
\bibitem{Nielsen1} A. B. Nielsen and J. H. Yoon, Classical Quantum Gravity {\bf 25}, 085010 (2008).
\bibitem{Nielsen2} M. Pielahn, G. Kunstatter and A. B. Nielsen Phys. Rev. D {\bf 84}, 104008 (2011).

\end{thebibliography}
\end{document}